\documentclass{svproc}
\usepackage{graphicx}
\usepackage{comment}
\usepackage{url}

\begin{document}
\mainmatter              
\title{Weighted Network Analysis of Biologically Relevant Chemical Spaces}
\titlerunning{Bioactive Compounds Networks}  
%
\author{Mariko I. Ito \and Takaaki Ohnishi}
\authorrunning{Ito and Ohnishi} 
%
%
\institute{The University of Tokyo, Bunkyo-ku Tokyo, Japan,\\
\email{marikoito.rnu1@gmail.com}\\ 
}

\maketitle              

\begin{abstract}
In cheminformatics, network representations of the space of compounds have been suggested extensively. Among these, the threshold-network consists of nodes representing molecules. In this network representation, two molecules are connected by a link if the Tanimoto coefficient, a similarity measure, between them exceeds a preset threshold. However, the topology of the threshold-network is affected significantly by the preset threshold. In this study, we collected the data of biologically relevant compounds and bioactivities. We defined the weighted network where the weight of each link between the nodes equals the Tanimoto coefficient between the bioactive compounds (nodes) without using the threshold. We investigated the relationship between the strength of the link connection and the bioactivity closeness in the weighted networks. We found that compounds with significantly high or low bioactivity have a stronger connection than those in the overall network. 
\end{abstract}
\section{Introduction \label{sec:intro}}
The chemical space is an abstract concept but is roughly defined as a set of all possible molecules~\cite{opassi2018hitchhiker,vogt2016lessons}.
In cheminformatics, the central idea that structurally similar compounds tend to share the similar chemical properties is called the {\it similarity property principle}~\cite{vogt2018progress}.
Based on this idea, the calculation of structural similarity is performed for various purposes from drug discovery to retrosynthetic analysis~\cite{kunimoto2017tracing,kunimoto2018combining,coley2017computer}.
In drug discovery, biologically relevant chemical spaces are primarily explored. 
Compounds exhibit biological activity in this space.
For example, ligand is a compound that binds to a receptor (the target) and inhibits biological response~\cite{dobson2004chemical}.
In these circumstances, it has been extensively investigated whether compounds with a similar structure share similar bioactivity~\cite{opassi2018hitchhiker}.

In networks, edges represent various kinds of relationships such as interaction, social influence, and correlation between two nodes~\cite{newman2003structure,albert2002statistical,ohnishi2010network,ito2018emergence,bazzi2016community}.
Investigating the topology of such networks affords a global view of how they are related to each other. 
For example, through community detection on networks, we can extract the groups of nodes each of which are densely connected. Nodes in a community can be regarded as those that are particularly interacting~\cite{mizokami2017revealing}, and having a similar feature or role~\cite{fortunato2010community}.
`Being similar' is a kind of relationship.
Previous studies have suggested certain network representations of biologically relevant chemical spaces, where each node represents a compound 
and each link shows the similarity relationship between the compounds~\cite{vogt2016lessons,vogt2018progress,kunimoto2017tracing,kunimoto2018combining,wu2016design,zhang2015design,zwierzyna2015design}.
They investigated the topological features of the chemical subspace and examined how molecules with certain bioactivity are distributed among the network.
In these studies, community detection was performed as well.
Through community detection on such networks, we obtain groups containing nodes with a similar chemical structure.

Network representation of a chemical space was performed using the circular fingerprint technique and Tanimoto coefficient~\cite{zwierzyna2015design}.
In circular fingerprint representation, a molecule is often represented by a vector called a fingerprint.
In the vector, each index denotes a certain chemical substructure, and the entry denotes the count of the molecule substructure corresponding to the index.
The Tanimoto coefficient is the most popular similarity measure between two molecules~\cite{lipkus1999proof}. 
It takes a value from 0 to 1, and equals 1 if two molecules are the same.
In previous studies regarding network representation based on the Tanimoto coefficient,
two nodes were assumed to be connected by a link 
if the Tanimoto coefficient between them exceeded a preset threshold.
In these studies, a threshold-dependent unweighted network is defined, known as the `threshold-network'.
Furthermore, the value of the preset threshold was tuned such that the edge density was approximately 0.025. 
Consequently, a well-resolved community structure was obtained.
Although the evaluation was not performed in detail, the visualized network demonstrated that compounds with similar bioactivity tend to form a community~\cite{zwierzyna2015design}.
However, the topology of the threshold-network is affected significantly by the preset threshold.
A point of concern is that the threshold-network constructed by an artificially preset threshold cannot capture the structure of the chemical space.
While constructing the threshold-network, the structural information of the chemical subspace should be reduced significantly.

Hence, in the present study, we analyze the weighted network of biologically relevant chemical spaces as follows.
Instead of applying a preset threshold to determine the existence of a link, 
we assume that two nodes (molecules) are connected by a link whose weight is the similarity between them. 
In particular, we are interested in discovering whether the weighted network topology can facilitate the investigation of compounds with high bioactivity.
We evaluate the community structure on the weighted networks and discuss whether nodes that are strongly connected to each other share a similar activity.

\section{Materials and Methods \label{sec:method}}
To investigate the structure of biologically relevant chemical spaces, we collected data from ChEMBL (version 25), an open bioactivity database~\cite{bento2014chembl}. 
We selected 19 targets based on a previous study~\cite{zwierzyna2015design}, as shown in Table~\ref{tab:data}. 
For each target, we extracted the data of compounds whose potency has been tested against the target by the measure of Ki, a kind of bioactivity.
We regard the pChEMBL value of the compound as an indicator of its bioactivity~\cite{bento2014chembl,steinmetz2015screening}.
The larger the pChEMBL value, the stronger is the bioactivity against the target~\cite{steinmetz2015screening}.
The number of compounds corresponding to each target is shown in Table~\ref{tab:data}.
Subsequently, we obtained the ``Morgan fingerprints'' (circular fingerprints) of these compounds using RDKit, an open-source toolkit for cheminformatics.
The circular fingerprint we used in this study is Morgan fingerprint. Each fingerprint is a 2048-dimensional vector, in which the entry in each index is an integer.
We calculated the Tanimoto coefficient for all pairs of compounds that correspond to each target.
For two molecules that have fingerprints $\mathbf{x}_i$ and $\mathbf{x}_j$,
the Tanimoto coefficient $T_{ij}$~\cite{lipkus1999proof} is calculated as
\begin{equation}
    T_{ij} = 
    \displaystyle
    \frac{\mathbf{x}_i\cdot\mathbf{x}_j}
    {|\mathbf{x}_i|^2 + |\mathbf{x}_j|^2 - \mathbf{x}_i\cdot\mathbf{x}_j}.
\end{equation}

Subsequently, the similarity matrix $T$, in which the $(i, j)$ entry is the similarity (Tanimoto coefficient) between molecules $i$ and $j$, is constructed for each target.
We considered the weighted network for each target by regarding the similarity matrix as the adjacency matrix.

In weighted network analysis, not only the number of links connected to the node, but also the sum of the weight of those links should be considered~\cite{barrat2004architecture}. 
The former is the degree of the node and the latter is its {\it strength}~\cite{lu2016vital}.
As the Tanimoto coefficient does not vanish in the case of almost all pairs of compounds, the weighted networks are almost complete and the degrees of nodes do not vary. 
Therefore, we examined how strength is distributed in each weighted network.
To examine whether the weighted networks exhibit community structure, 
we applied Louvain heuristics, an algorithm used to obtain a graph partition that (locally) optimizes the modularity $Q$~\cite{blondel2008fast}.
In the case of weighted networks, the modularity $Q$~\cite{lu2016vital,fortunato2010community} is defined as
\begin{equation}
    Q = \displaystyle\frac{1}{2m}
    \sum_{i, j} M_{ij}\delta(c(i), c(j)),
\end{equation}
where
\begin{equation}
    M_{ij} := T_{ij} - \displaystyle\frac{s_i s_j}{2m}.
    \label{eq:mij}
\end{equation}
The strength of node $i$, $\sum_j T_{ij}$, is denoted by $s_i$.
The sum of all weights $\sum_{i,j} T_{i j}$ is $2m$ 
and $c(i)$ denotes the community to which node $i$ belongs.
Kronecker's delta is denoted by $\delta$; therefore, $\delta (x, y)$ equals 1 (0) when $x=y$ ($x\neq y$). 
Regarding $M_{ij}$, the second term on the right side of Eq.~(\ref{eq:mij}) represents the expected strength of the link between nodes $i$ and $j$ in the null-network, which is random except that it has the same strength distribution as the focal network~\cite{fortunato2010community}.
Therefore, $M_{ij}$ represents how strongly nodes $i$ and $j$ are connected compared to the null-model.

\begin{table}
\caption{{\bf Examined compounds.} In the first column, Target ID means the ChEMBL ID assigned to each target in the ChEMBL database. The second column shows the name of the targets---Hs, Rn, and Cp represent {\it Homo sapiens}, {\it Cavia Porcellus}, and {\it Rattus norvegicus}, respectively.  In the third column, the number of compounds that correspond to the target is shown. In the fourth and fifth columns, the mean and standard deviation of the node strength are shown, respectively. The sixth and seventh columns show the number of communities and the modularity $Q$ in the resulted graph partition by the community detection, respectively.}
\begin{center}
\begin{tabular}{llrrrrr}
\hline
Target ID \quad & Target name & ~Size & ~Mean & ~Std & ~~Com & ~~~~~~~$Q$\\
\hline
255 & Adenosine A2b receptor (Hs) & 1575 & 723 & 130 & 3 & 0.060\\
3242 & Carbonic anhydrase XII (Hs) & 2392  & 863 & 262 & 3 & 0.051\\
269 & Delta opioid receptor (Rn) & 1577 & 719 & 113 & 4 & 0.081\\
219 & Dopamine D4 receptor (Hs) & 2138 & 1087 & 183 & 4 & 0.031\\
238 & Dopamine transporter (Hs) & 1406 & 528 & 86 & 4 & 0.065\\ 
65338 & Dopamine transporter (Rn) & 1624 & 723 & 105 & 3 & 0.074\\
339 & Dopamine D2 receptor (Rn) & 2555 & 1119 & 171 & 3 & 0.045\\
4124 & Histamine H3 receptor (Rn) & 1591 & 637 & 135 & 4 & 0.075\\
344 & Melanin-concentrating & 1430 & 771 & 68 & 4 & 0.046\\
    & hormone receptor 1 (Hs) & & & & \\
270 & Mu opioid receptor (Rn) & 2318 & 984 & 178 & 4 & 0.091\\
4354 & Mu opinion receptor (Cp) & 654 & 266 & 51 & 3 & 0.078\\
2014 & Nociceptin receptor (Hs) & 1105 & 519 & 97 & 4 & 0.070\\
2001 & Purinergic receptor P2Y12 (Hs) & 584 & 400 & 74 & 4 & 0.029\\
225 & Serotonin 2c (5-HT2c) receptor (Hs) & 1980 & 785 & 132 & 4 & 0.049\\
273 & Serotonin 1a (5-HT1a) receptor (Rn) & 3370 & 1469 & 249 & 3 & 0.052\\
322 & Serotonin 2a (5-HT2a) receptor (Rn) & 3076 & 1278 & 215 & 4 & 0.067\\
1833 & Serotonin 2b (5-HT2b) receptor (Hs) & 1121 & 421 & 74 & 5 & 0.058\\
3155 & Serotonin 7 (5-HT7) receptor (Hs) & 1569 & 775 & 129 & 3 & 0.035\\
4153 & Sigma-1 receptor (Cp) & 1617 & 717 & 123 & 3 & 0.048\\[2pt]
\hline
\end{tabular}
\end{center}
\label{tab:data}
\end{table}

\section{Results \label{sec:result}}
\begin{figure}
\begin{center}
\includegraphics[width = 12cm]{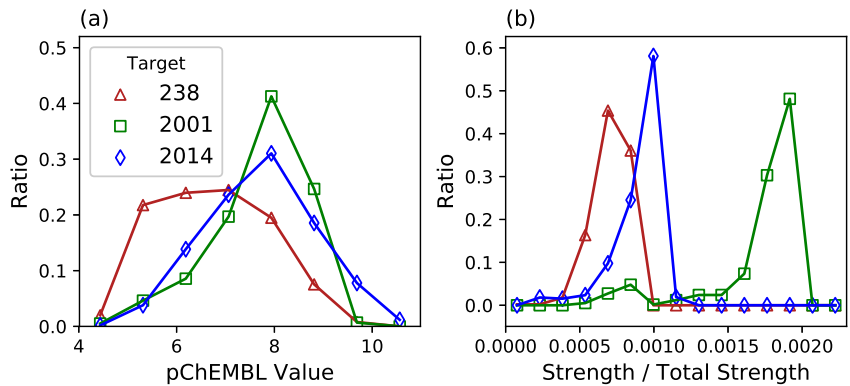}
\caption{\label{fig:deg_pchembl} {\bf Histogram of pChEMBL values and strength.} (a) Histograms of pChEMBL values in networks of targets 238, 2001, and 2014. (b) Histograms of strength normalized by total strength in networks of targets 238, 2001, and 2014. Markers are same as in (a).}
\end{center}
\end{figure}

First, we show the histogram of pChEMBL value in each compound set corresponding to each target.
For the three examples of networks for targets 238, 2001, and 2014, the histograms of pChEMBL values are shown in Fig.~\ref{fig:deg_pchembl}(a).
Few compounds have an extremely small or large value.
We also observed a similar tendency in the case of other targets.

Subsequently, we examined the strength of the nodes in the weighted networks.
In Table~\ref{tab:data}, we show the mean and standard deviation of the node strength in the weighted network for each target.
Fig.~\ref{fig:deg_pchembl}(b) shows three histograms of the strength normalized by the total strength, $s_i / \sum_{i,j} T_{ij}$, in each network.
As shown, many nodes share a similar strength, while a few nodes exhibit small strength.
No node exhibited extremely large strength in all networks.

Finally, we investigated whether nodes connected with high similarity tend to share similar bioactivity.
As explained in Sect.~\ref{sec:method}, we performed community detection.
The number of communities and the modularity $Q$ of the graph partition resulted from the Louvain heuristics is shown for each target in Table~\ref{tab:data}.
The values of modularity are low, 
and the community structure in each network is weak in general.

\begin{figure}
\begin{center}
\includegraphics[width = 12cm]{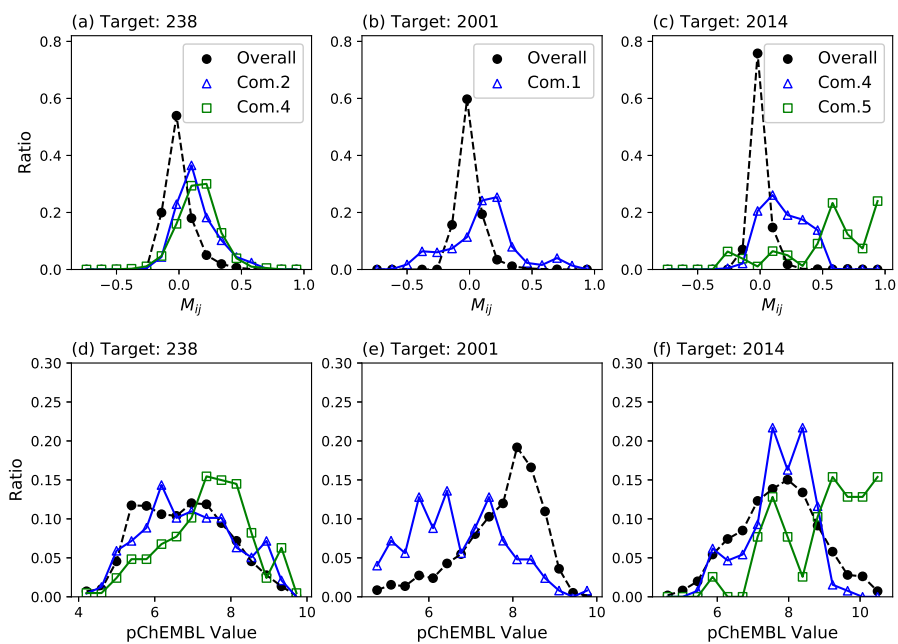}
\caption{\label{fig:community} {\bf Histograms of $M_{ij}$ and pChEMBL values in communities.} Histograms of $M_{ij}$ in the overall network and in each community in the cases of targets 238 (a), 2001 (b), and 2014 (c). Histograms of pChEMBL values in the overall network and in each community in the cases of targets 238 (d), 2001 (e), and 2014 (f). In these figures, only communities with $Q_C > 0.2$ and size exceeding 20 are shown.}
\end{center}
\end{figure}

We further inspected the community structure obtained by this community detection.
Some detected communities were not connected sufficiently; as such, they could not be called `communities'. 
Therefore, we extracted communities that could be regarded as connected strongly.
For detected community $C$, we defined the extent to which the nodes are strongly connected within $C$, $Q_C$, as
\begin{equation}
    Q_C = \displaystyle\frac{1}{\sum_{i, j\in C} T_{ij}}
    \sum_{i, j\in C}M_{ij},
\end{equation}
where $ \sum_{i, j\in C}$ denotes the sum of all pairs of nodes within community $C$.
Therefore, the modularity $Q$ equals $\sum_{C}Q_C$, and $Q_C$ can measure how strongly links within $C$ are connected without considering other communities.
In Figs.~\ref{fig:community}(a)--(c), we show the histograms of $M_{ij}$ for all pairs of nodes within each community and those in the overall network, for three targets.
In these figures, only communities with $Q_C$ exceeding 0.2 are shown.
Therefore, these communities have larger values of $M_{ij}$ than the overall network.

\begin{figure}
\begin{center}
\includegraphics[width = 8cm]{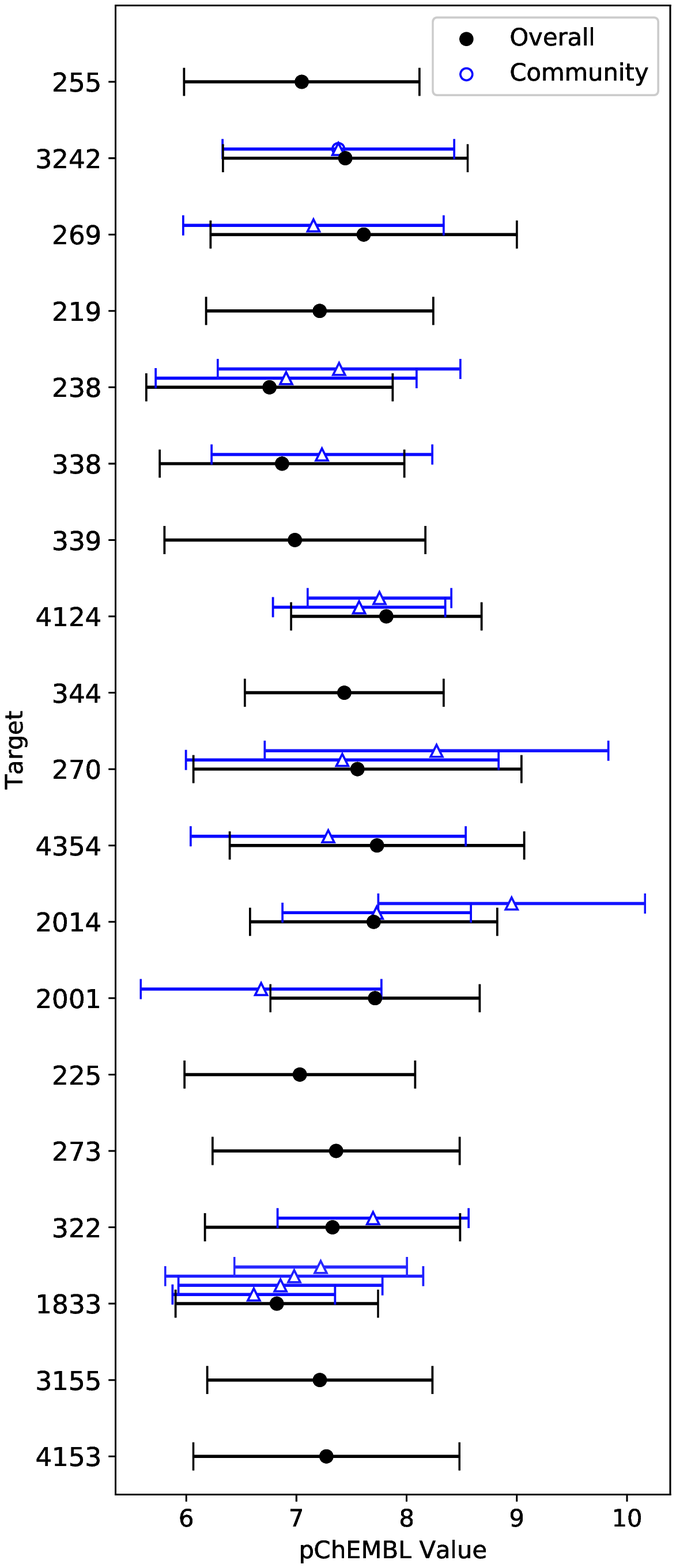}
\caption{\label{fig:all_community} {\bf Mean pChEMBL value in each community.} For each target (ordinate), the mean pChEMBL value in the overall network is shown by a circle. For communities with $Q_C>0.2$ and size exceeding 20, the mean pChEMBL value in each community is shown by a triangle above that of the overall network. The error-bar represents the standard deviation of pChEMBL values.}
\end{center}
\end{figure}

On the other hand, Figs.~\ref{fig:community}(d)--(f) show the histogram of pChEMBL values of nodes in each community included in Figs.~\ref{fig:community}(a)--(c) and the overall network.
In the case of target 238,
the histogram of pChEMBL value for community 4 shifts to the right side compared to the overall network (Fig.~\ref{fig:community}(d)). 
Community 5 in the network for target 2014 exhibits the same feature as well (Fig.~\ref{fig:community}(f)).
Conversely, Community 1 in the network for target 2001 comprise nodes with lower pChEMBL values than those in the overall network.
In Fig.~\ref{fig:all_community}, 
for all targets, we show the mean of pChEMBL values in each community that satisfies $Q_C > 0.2$ and consists of more than 20 nodes.
Each error bar shows the standard deviation.
In some communities, the mean pChEMBL value is located far from that of the overall network.
However, in most cases, this value is within the standard deviation range of that of the overall network.

In summary, although the whole community structure is weak, we observed some communities in which the nodes are connected with a large weight.
In some of them, the distribution of pChEMBL value is biased compared to that of the overall network.
This suggests that certain sets of compounds are similar to each other and share stronger/weaker bioactivity against the target than the compounds in general. 

\begin{figure}
\begin{center}
\includegraphics[width = 12cm]{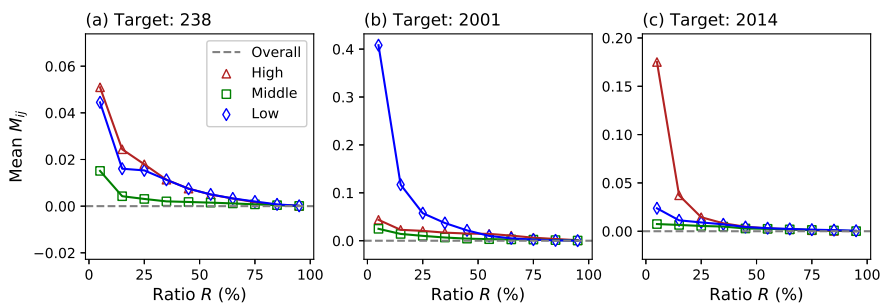}
\caption{\label{fig:top} {\bf Mean $M_{ij}$ as a function of the ratio $R$.} Mean $M_{ij}$ of all links that connect nodes with high (more than ($100-R$)-th percentile)/intermediate (from ($50-R/2$)-th to ($50+R/2$)-th percentile)/low (less than $R$-th percentile) pChEMBL values versus the ratio $R$, in the cases of targets 238 (a), 2001 (b), and 2014 (c).}
\end{center}
\end{figure}

Subsequently, we investigated whether nodes with particularly high (low) pChEMBL values are connected to each other with a large weight.
First, we collected the $\lfloor 0.01RN\rfloor$ nodes whose pChEMBL values exceeded the ($100-R$)-th percentile, where $N$ is the number of nodes in the network.
Second, we calculated the mean of $M_{ij}$ for all pairs of the $\lfloor 0.01RN\rfloor$ nodes. 
Similarly, we calculated the mean of $M_{ij}$ for nodes with low pChEMBL values (lower than the $R$-th percentile), and intermediate pChEMBL values (ranging from the ($50-R/2$)-th to ($50+R/2$)-th percentile).
The results for these three cases are presented in Figs.~\ref{fig:top}(a)--(c), where the horizontal axis represents the ratio $R$ and the vertical axis the mean of $M_{ij}$.
The mean $M_{ij}$ exceeded 0 when the ratio $R$ is small in the cases of high and low pChEMBL values. 
The mean $M_{ij}$ also exceeds 0 for a small ratio $R$ in the case of intermediate pChEMBL values, but it is much lower than the means in the other cases.
The mean $M_{ij}$ decreases with the ratio and approaches the mean of the overall network, which approximately equals 0.
Although Figs.~\ref{fig:top}(a)--(c) show only the targets 238, 2001, and 2014, we observed the same tendency in all other targets.
Therefore, the sets of nodes with high/low pChEMBL values in particular are connected with stronger weights than the overall network.
Figs.~\ref{fig:top}(a)--(c) show some consistency with Figs.~\ref{fig:community}(d)--(f).
For target 2001, the nodes with low pChEMBL values are connected strongly (Fig.~\ref{fig:top}(b)) 
and some of them are detected as those included in Community 1 sharing a low pChEMBL value (Fig.~\ref{fig:community}(e)).
Additionally, consistency is shown between Community 5 in the network of target 2014 (Fig.~\ref{fig:community}(f)) and the set of high pChEMBL values (Fig.~\ref{fig:top}(c)).

\section{Discussion \label{sec:discussion}}
In this study, we investigated the structure of biologically relevant compounds that share the same target.
Previous studies have suggested the network representations of the structure.
In the network representation of these compounds, a node represents a compound, and a link between two nodes is drawn if the similarity between them exceeds a preset threshold.
The topology of these threshold-networks greatly depends on the preset threshold.
Therefore, to understand the true nature of the structure, we considered the weighted network, where the weight of each link is the similarity between connecting compounds.

For each target, the corresponding weighted network showed a homogeneous structure, 
which comprises a rare node exhibiting extremely strong bioactivity, or that connecting to other nodes with extreme strength.
This homogeneity is attributable to the sample bias---the compounds in each network are those sharing the same target.

In cheminformatics, the question of whether compounds that are structurally similar share similar chemical properties needs to be elucidated~\cite{opassi2018hitchhiker}.
We performed community detection on the weighted networks to investigate whether strongly connected nodes exhibit similar bioactivity.
We found that, in general, the community structure was weak in all the weighted networks.
However, we observed that nodes with high/low bioactivity against the target were connected strongly to each other compared to the nodes of the overall network.
Some detected communities reflected this tendency and each of their nodes exhibited a different range of pChEMBL values compared to those in the overall network.
As a practical application, such communities can help us predict whether a novel compound exhibits high bioactivity.
If the novel compound is structurally similar to compounds in a community sharing high bioactivity, we can expect it to exhibit high bioactivity.    
Such a prediction is useful in drug discovery.

In a previous study concerning the threshold-network, community detection was performed using modularity as the quality function of graph partition~\cite{zwierzyna2015design}.
The values of modularity were much greater than those in our study and a well-resolved community structures were obtained.
In fact, the threshold was set to obtain high modularity without resulting in an extremely sparse community structure.
The threshold was set based on network visualization, which appears to be intuitive.
The weight of a link should be determined mathematically considering observations from a community structure; this is aimed to be explored in future works.
For example, a definition of the weight link may be represented as a sigmoid curve
\begin{equation}
    f(x) = \displaystyle\left[\frac{1}{1+\left(\frac{1}{x^\alpha} - 1\right)^\beta}\right]^\gamma,
\end{equation}
where $x (\in\left[0,1\right])$ is the similarity.
As $\alpha$ approaches infinity, $f(x)$ approaches 1 (0) when $x>2^{-1/\alpha}$ ($x<2^{-1/\alpha}$).  
This limit of $\beta\rightarrow \infty$ corresponds to the construction of the threshold-network, in which the preset threshold is $2^{-1/\alpha}$.
On the other hand, setting the weight of each link in our study corresponds to a limiting function $f(x) = x$, which is obtained when $\alpha$, $\beta$ and $\gamma$ equal 1. 
In future works, the optimization of parameters $\alpha$, $\beta$ and $\gamma$ should be investigated.
Accordingly, it will be challenging to evaluate the efficacy of the detected community structure considering the structural similarity, bioactivity distribution, and application, for example, to drug discovery.

Finally, the weighted network representation with other definitions of similarity should be considered.
Although the Tanimoto coefficient is a popular similarity measure, 
it measures the global similarity of compounds, which is sometimes disadvantageous.
The bioactivity of a compound often depends on the structure of a certain part of the compound.
In some studies, networks were considered and evaluated based on other types of similarity~\cite{vogt2016lessons,kunimoto2018combining,wu2016design}.
Weighted networks with those similarity measures have not been investigated yet.
We expect that weighted network analysis with similarity measures other than the Tanimoto coefficient can promote a better understanding of the structure--activity relationship in a biologically relevant chemical space.

\section*{Acknowledgement}
This work was supported by the JSPS Grant-in-Aid for Scientific Research on Innovative Areas: 17H06468.

\bibliographystyle{ieeetr}

\begin{thebibliography}{10}

\bibitem{opassi2018hitchhiker}
G.~Opassi, A.~Ges{\`u}, and A.~Massarotti, ``The hitchhiker’s guide to the
  chemical-biological galaxy,'' {\em Drug discovery today}, vol.~23, no.~3,
  pp.~565--574, 2018.

\bibitem{vogt2016lessons}
M.~Vogt, D.~Stumpfe, G.~M. Maggiora, and J.~Bajorath, ``Lessons learned from
  the design of chemical space networks and opportunities for new
  applications,'' {\em Journal of computer-aided molecular design}, vol.~30,
  no.~3, pp.~191--208, 2016.

\bibitem{vogt2018progress}
M.~Vogt, ``Progress with modeling activity landscapes in drug discovery,'' {\em
  Expert opinion on drug discovery}, vol.~13, no.~7, pp.~605--615, 2018.

\bibitem{kunimoto2017tracing}
R.~Kunimoto, M.~Vogt, and J.~Bajorath, ``Tracing compound pathways using
  chemical space networks,'' {\em MedChemComm}, vol.~8, no.~2, pp.~376--384,
  2017.

\bibitem{kunimoto2018combining}
R.~Kunimoto and J.~Bajorath, ``Combining similarity searching and network
  analysis for the identification of active compounds,'' {\em ACS omega},
  vol.~3, no.~4, pp.~3768--3777, 2018.

\bibitem{coley2017computer}
C.~W. Coley, L.~Rogers, W.~H. Green, and K.~F. Jensen, ``Computer-assisted
  retrosynthesis based on molecular similarity,'' {\em ACS central science},
  vol.~3, no.~12, pp.~1237--1245, 2017.

\bibitem{dobson2004chemical}
C.~M. Dobson, ``Chemical space and biology,'' 2004.

\bibitem{newman2003structure}
M.~E. Newman, ``The structure and function of complex networks,'' {\em SIAM
  review}, vol.~45, no.~2, pp.~167--256, 2003.

\bibitem{albert2002statistical}
R.~Albert and A.-L. Barab{\'a}si, ``Statistical mechanics of complex
  networks,'' {\em Reviews of modern physics}, vol.~74, no.~1, p.~47, 2002.

\bibitem{ohnishi2010network}
T.~Ohnishi, H.~Takayasu, and M.~Takayasu, ``Network motifs in an inter-firm
  network,'' {\em Journal of Economic Interaction and Coordination}, vol.~5,
  no.~2, pp.~171--180, 2010.

\bibitem{ito2018emergence}
M.~I. Ito, H.~Ohtsuki, and A.~Sasaki, ``Emergence of opinion leaders in
  reference networks,'' {\em PloS one}, vol.~13, no.~3, p.~e0193983, 2018.

\bibitem{bazzi2016community}
M.~Bazzi, M.~A. Porter, S.~Williams, M.~McDonald, D.~J. Fenn, and S.~D.
  Howison, ``Community detection in temporal multilayer networks, with an
  application to correlation networks,'' {\em Multiscale Modeling \&
  Simulation}, vol.~14, no.~1, pp.~1--41, 2016.

\bibitem{mizokami2017revealing}
C.~Mizokami and T.~Ohnishi, ``Revealing persistent structure of international
  trade by nonnegative matrix factorization,'' in {\em International Conference
  on Complex Networks and their Applications}, pp.~1088--1099, Springer, 2017.

\bibitem{fortunato2010community}
S.~Fortunato, ``Community detection in graphs,'' {\em Physics reports},
  vol.~486, no.~3-5, pp.~75--174, 2010.

\bibitem{wu2016design}
M.~Wu, M.~Vogt, G.~M. Maggiora, and J.~Bajorath, ``Design of chemical space
  networks on the basis of tversky similarity,'' {\em Journal of computer-aided
  molecular design}, vol.~30, no.~1, pp.~1--12, 2016.

\bibitem{zhang2015design}
B.~Zhang, M.~Vogt, G.~M. Maggiora, and J.~Bajorath, ``Design of chemical space
  networks using a tanimoto similarity variant based upon maximum common
  substructures,'' {\em Journal of computer-aided molecular design}, vol.~29,
  no.~10, pp.~937--950, 2015.

\bibitem{zwierzyna2015design}
M.~Zwierzyna, M.~Vogt, G.~M. Maggiora, and J.~Bajorath, ``Design and
  characterization of chemical space networks for different compound data
  sets,'' {\em Journal of computer-aided molecular design}, vol.~29, no.~2,
  pp.~113--125, 2015.

\bibitem{lipkus1999proof}
A.~H. Lipkus, ``A proof of the triangle inequality for the tanimoto distance,''
  {\em Journal of Mathematical Chemistry}, vol.~26, no.~1-3, pp.~263--265,
  1999.

\bibitem{bento2014chembl}
A.~P. Bento, A.~Gaulton, A.~Hersey, L.~J. Bellis, J.~Chambers, M.~Davies, F.~A.
  Kr{\"u}ger, Y.~Light, L.~Mak, S.~McGlinchey, {\em et~al.}, ``The chembl
  bioactivity database: an update,'' {\em Nucleic acids research}, vol.~42,
  no.~D1, pp.~D1083--D1090, 2014.

\bibitem{steinmetz2015screening}
F.~P. Steinmetz, C.~L. Mellor, T.~Meinl, and M.~T. Cronin, ``Screening
  chemicals for receptor-mediated toxicological and pharmacological endpoints:
  Using public data to build screening tools within a knime workflow,'' {\em
  Molecular informatics}, vol.~34, no.~2-3, pp.~171--178, 2015.

\bibitem{barrat2004architecture}
A.~Barrat, M.~Barthelemy, R.~Pastor-Satorras, and A.~Vespignani, ``The
  architecture of complex weighted networks,'' {\em Proceedings of the national
  academy of sciences}, vol.~101, no.~11, pp.~3747--3752, 2004.

\bibitem{lu2016vital}
L.~L{\"u}, D.~Chen, X.-L. Ren, Q.-M. Zhang, Y.-C. Zhang, and T.~Zhou, ``Vital
  nodes identification in complex networks,'' {\em Physics Reports}, vol.~650,
  pp.~1--63, 2016.

\bibitem{blondel2008fast}
V.~D. Blondel, J.-L. Guillaume, R.~Lambiotte, and E.~Lefebvre, ``Fast unfolding
  of communities in large networks,'' {\em Journal of statistical mechanics:
  theory and experiment}, vol.~2008, no.~10, p.~P10008, 2008.

\end{thebibliography}

\end{document}